\documentstyle[art10,hhep,epsf]{article}
\begin{document}
\cimo \setcounter{page}{1} \thispagestyle{empty} \hskip -15pt  \noindent
{\Large\bf
Propagation of Strange Quark Matter in the \\ 
Atmosphere
}\\[3mm] 
\def\rightmark{Propagation of Strange Quark Matter...}\def\leftmark{G.Wilk, et al.}
\hspace*{6.327mm}\begin{minipage}[t]{12.0213cm}{\large\lineskip .75em
Grzegorz Wilk$^{1}$ and Zbigniew W\l odarczyk$^{2}$
}\\[2.812mm] 
\hspace*{-8pt}$^1$ Soltan Institute for Nuclear Studies, \\
00-681 Warsaw, Poland\\[0.2ex]
\hspace*{-8pt}$^2$ Institute of Physics, Pedagogical University, \\
25-509 Kielce, Poland
\\[4.218mm]{\it
As for 30 June 1996.
}\\[5.624mm]\noindent
{\bf Abstract.} We propose model of propagation of lumps of Strange
Quark Matter (strangelets) through the atmosphere, which accounts for
their apparent strong penetrability and normal nuclear-type sizes at
the same time. The mass spectrum of strangelets reaching the Earth
predicted by this model agrees very well with the existing data on
the abundance of different elemets in the Universe.
\end{minipage}

\section{Introduction}
 
In astrophysical literature one finds a number of phenomena which can
be regarded as possible manifestation of the {\it strange quark
matter} (SQM), extremaly interesting possibility of new stable form
of matter \cite{SQM,AO,FB,ALOCK}. There were attempts to find lumps
of SQM called {\it strangelets} in the terrestial experiments devoted
to search for Quark Gluon Plasma but so far without success
\cite{SEARCH} (what can be interpreted that no QGP was formed so far 
in these experiments \cite{SPIELES}). The strangelets were search for 
also in cosmic ray (CR) events \cite{SAITO,PRICE,ICH,JACEE,CAP}, the
most famous candidate being the so called {\it Centauro} events
\cite{LATTES}. In CR one either deals with strangelets formed by
some astrophysical mechanisms \cite{SQM,HAENSEL} or one witnesses
their production proceeding in collisions of the original CR flux
with atmospheric nuclei \cite{PANAG}.  

{\it Centauro} events are good example of problem being the topic of
our presentation. They were observed very deep in the atmosphere (at
$\sim 500$ g/cm$^2$) and if they would be really caused by
strangelets \cite{PANAG}, it would require unusual penetrability of
such objects implying their small geometrical sizes (much smaller
than the typical nuclear size). Similar conclusions were also reached
in some CR experiments \cite{JACEE,CAP}. We present an alternative
approach to the propagation of strangelets through the atmosphere in
which we argue that they have most probably normal nuclear-type sizes
and similar propagation properties. We then demonstrate how such
strangelets can penetrate deeply in the atmosphere producing effects
ascribed before to higly penetrating objects. Finally, we estimate
the initial flux of strangelets entering atmosphere emerging from our
picture and compare it to the known abundances of normal nuclei in
Universe.

\section{Strangelets}

Typical SQM consists of roughly equal number of $u$, $d$ and $s$
quarks and it has been found to be the true ground state of QCD
\cite{SQM,AO,FB}, i.e., it is absolutely stable at high mass numbers
$A$. However, any SQM produced at very early stage of the history
of Universe would have evaporated long time ago \cite{ALOCK}. On the
other hand, there are places where  SQM may still exist at present
\cite{SQM,ALOCK}. It is probably continously produced in neutron
stars with a superdense quark surface and in quarks stars with a thin
nucleon envelope \cite{HAENSEL,SQM}. Collisions of such objects could
therefore produce small lumps of SQM, strangelets with $10^2 < A <
10^6$, permeating the Galaxy and possibly reaching also the Earth
(i.e., {\it a priori} being detectable here). 

The practical measure of the stability of strangelet is provided by
the so called separation energy $dE/dA$, i.e., energy, which is
required to remove a single baryon from a strangelet. For example, if
$dE/dA > m_N$ then strangelet can evaporate neutrons from its
surface. This energy depends, among other things, on the size of
strangelet, i.e., on its mass number $A$ \cite{FB}. There exists
therefore some critical size $A = A_{crit}~(\sim 300~-~400$ depending
on the parameters used) such that for $A < A_{crit}$ strangelets are
unstable against neutron emission \cite{FB}.  

Anomalous massive particles, which can be interpreted as strangelets,
have been apparently observed in three independent CR experiments:\\
\noindent
{\bf (i)}~~~In counter experiment devoted to study primary CR nuclei
two anomalous events have been observed \cite{SAITO}. They are
consistent with values of charge $Z \simeq 14$ and of mass numbers
$A \simeq 350$ and $\simeq 450$ and cannot be accounted for by the
conventional background. Such values of $Z$ and $A$ are fully
consistent with the theoretical estimation for ratio $Z/A$ in SQM 
\cite{KASUYA}.\\
\noindent
{\bf (ii)}~~~The so called Price's event \cite{PRICE} with $Z \simeq
46$ and $A > 1000$, regarded previously as possible candidate for
magnetic monopole, turned out to be fully consistent with the above 
ratio $Z/A$ for SQM \cite{S}.\\
\noindent
{\bf (iii)}~~~The so called Exotic Track with $Z \simeq 20$ and $A
\simeq 460$ has been reported in \cite{ICH}. The name comes from the
fact that although it was observed in emulsion chamber exposed to CR
on balloon at the atmospheric depth of only $11.7$ g/cm$^2$ its
arrival zenith angle of $87.4$ deg means that the projectile causing
that event traversed $\sim 200$ g/cm$^2$ of atmosphere (in contract
to events $(i)$ and $(ii)$ where the corresponding depths were of the
order of $5 - 15$ g/cm$^2$ of atmosphere only). 

The Exotic Track event motivated the (balloon born emulsion chamber)
JACEE \cite{JACEE} and Concorde aircraft \cite{CAP} experiments to
search for strangelets with such long mean free paths. In fact,
authors of \cite{JACEE,CAP} suggest that the interaction mean free
path for strangelets in atmosphere is of the order of $\lambda_{\rm
S} = 124$ g/cm$^2$ for $A = 100$ decreasing to $\lambda_{\rm S} = 59$
g/cm$^2$ only for $A = 1000$. These values are surprisingly close 
to  that for protons at comparable energies ($\lambda_{proton} =
60~-70~$  g/cm$^2$) and are much bigger than that for a normal
nucleus ($\lambda_{nucleus} \simeq 3.8$ g/cm$^2$ for $A = 100$)
\cite{F2}. It was then suggested in \cite{JACEE,CAP} that
strangelets should have geometrical radii much smaller than those of
the ordinary nuclei or, correspondingly, much smaller interaction
cross sections (in agreement with SQM interpretation of Centauro
events mentioned before \cite{PANAG}).

\vspace*{8.8cm}
\includegraphics{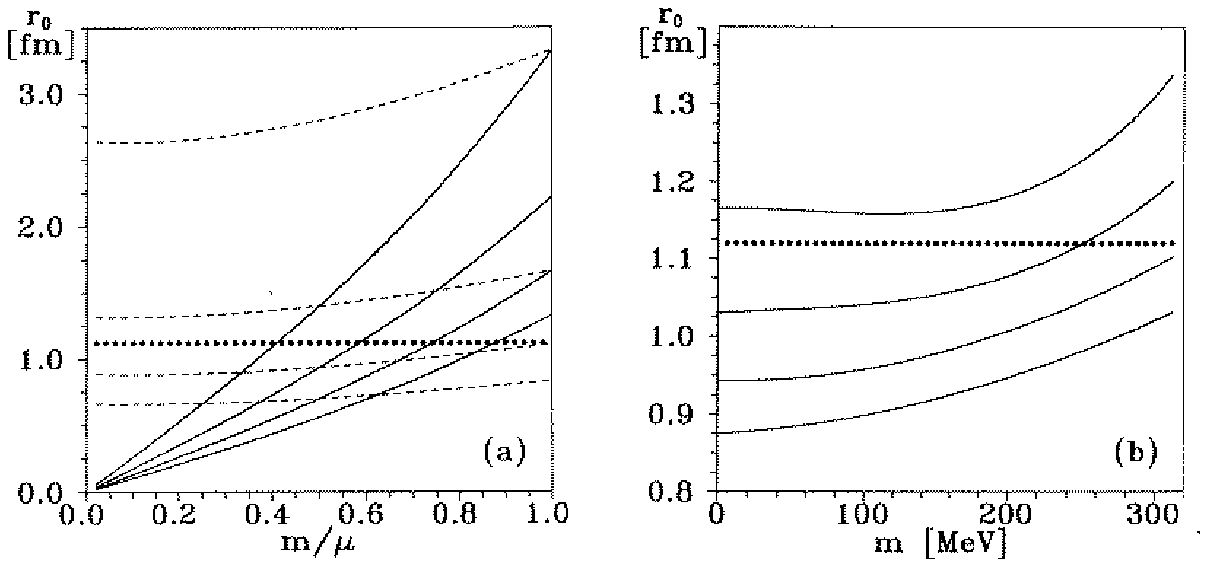}
\vskip -70pt
\begin{minipage}[t]{10cm}
\noindent \bf Fig. 1.  \rm
Dependence of the rescaled radius $r_0$ [fm] (eq.(\ref{eq:r0})):
$(a)$ on the ratio of $s$ quark mass $m$ to its chemical potential
$\mu$, $m/\mu$ (solid lines, read from top to bottom, correspond to
fixed $m = 100,~150,~200,~250$ MeV; dashed read from top to bottom
correspond to fixed $\mu = 100,~ 200,~300,~400$ MeV); $(b)$ on 
the strange quark mass $m$ for different values of $\alpha_c =
0.9,~0.6,~0.3,~0.0$ (from top to bottom, respectively). In both cases
dotted line shows $r_0 = 1.12$ fm corresponding to normal nuclear
density $\rho = 0.17~{\rm fm}^{-3}~=~(110~{\rm MeV})^3$. 
\end{minipage}
\vskip 4truemm
However, such expectations are not confirmed by calculations, cf.
Fig. 1. Let us consider a strangelet visualised as Fermi gas of $u$,
$d$ and $s$ quarks, with total mass number $A$, confined in spherical
volume $V \sim A$ of radius $R\, =\, r_0\,A^{1/3}$, where
the rescaled radius $r_0$ is determined by the number density of
strange matter, $n = A/V =  \left( \, n_u\, +\, n_d\, +\, n_s\,
\right)/3$, for which $r_0 = \left(3/4\pi n\right)^{1/3}$ . Here $n_i
\, =\, -\, \frac{\partial \Omega _i}{\partial \mu _i}$ and
thermodynamical potentials $\Omega_i(m_i,\mu_i)$ are related to
chemical potentials $\mu_i$ \cite{FB}. Because in our case all $\mu_i
\sim 300$ MeV, one can neglect the (current) masses of $u$ and $d$
quarks and keep only the $s$ quark mass $m$. Taking into account 
the QCD ${\cal O}(\alpha_c)$ corrections in
$\Omega_i(m_i,\mu_i,\alpha_c)$ and renormalizing them at $m_N/3 =
313$ MeV \cite{FB}, one arrives at 
\begin{equation}
r_0\, =\, \left\{\, \frac{3\, \pi}
           {2\, \left( 1 - \frac{2\alpha_c}{\pi}\right)
          \, \left[ \mu^3 + \left( \mu^2 - m^2\right) ^{3/2} \right] }
          \, \right\} ^{1/3} . \label{eq:r0}
\end{equation}
As one can see in Fig. $1a$ for $m \simeq 150$ MeV and $\mu \simeq
300$ MeV (the values commonly accepted for SQM \cite{SQM,FB}), the
values of $r_0$ of the strangelets are comparable with those for the
ordinary nuclear matter. Fig. $1b$ summarizes dependence of $r_0$ on
$\alpha_c$. As one can see QCD corrections lead to slight increase of
$r_0$. (It is worth to say that at this point we differ drastically
from previous works looking at strangelets from the QGP and Centauro
production perspective \cite{PANAG} where large value of $\mu$ was
the main reason for small values of $r_0$ and for all further claims
concerning large penetrability of strangelets. Such large values of
$\mu$ were justified there because strangelets were supposed to be
produced from the QGP phase formed in baryon rich environment whereas
in our case they are supposed to be stable object entering Earth
atmosphere as a part of CR radiation.). Fig. $1$ shows therefore that
the expected decrease of the radius of strangelet is nowhere as
dramatic as it has been estimated in Refs. \cite{JACEE,CAP}, i.e.,
the expected geometrical cross sections of SQM are not able to
explain alone the occurences of anomalous events detected deeply in
the atmosphere.   

\section{Possible scenario of propagation of strangelets through
the atmosphere} 

It does not mean, however, that some form of SQM does not reach 
Earth and penetrates deep in the atmosphere to be finally registered.
The apparent contradiction between its "normal" size and strong
penetrability can be resolved in very simple manner. In what follows
we shall propose the respective (speculative but plausible) scenario,
which can be summarized in the following way: {\it Strangelets
reaching so deeply into atmosphere are formed in many successive
interactions with air nuclei of much heavier lumps of SQM entering
our atmosphere}. In this way one can accomodate both the most
probably "normal" mean free paths for successive interactions and
final large penetrating depth. Such scenario is fully consistent with
all present and proposed experiments and could be additionally
checked only by measuring products of the intermediate collisions,
which so far is impossible.

To provide numerical estimations we shall limit ourselves to two
extreme pictures of collisions of strangelet of mass number $A$ with
air nucleus target of mass number $A_t$:\\
\noindent
{\bf 1.}~~~"Standard model": all quarks of $A_t$ which are located
in the geometrical intersection of two colliding projectiles are
involved and it is assumed that each quark from the target interacts
with only one quark from the strangelet; i.e., during interaction
$3A_t$ quarks from strangelet can be used up and after it the mass
number of strangelet is diminished to the value of $A-A_t$at most. In
this case simple analytical estimation is possible leading to the
total penetration depth of strangelet
\begin{equation}
\Lambda\, \simeq \, \frac{1}{3}\, \lambda_{NA_t}\, 
                     \left(\frac{A_0}{A_t}\right)^{1/3}\,
                     \left(1\, -\, \frac{A_{crit}}{A_0}\right)^2\, 
                     \left(4\, -\, \frac{A_{crit}}{A_0}\right) \, 
                     \cong \, \frac{4}{3}\, \lambda_{NA_t}\,
                   \left(\frac{A_0}{A_t}\right)^{1/3} .\label{eq:L}
\end{equation}
(where $\lambda_{NA_t}$ denotes the mean free path for $NA_t$
collisions). \\
\noindent
{\bf 2.} "Tube-like model": All quarks from both nuclei which are
in their geometrical intersection region participate in the
collision. This is an analogue of the so called tube model (TM) used
sometimes in nuclear collisions and after collision the atomic mass
of strangelet diminishes to the value equal to $A - A^{1/3}\cdot
A_t^{2/3}$ (this variant should be regarded as a limiting possibility
providing estimation of the maximal possible destruction of the
quarks in the strangelet). 

\vspace*{9cm}
\includegraphics{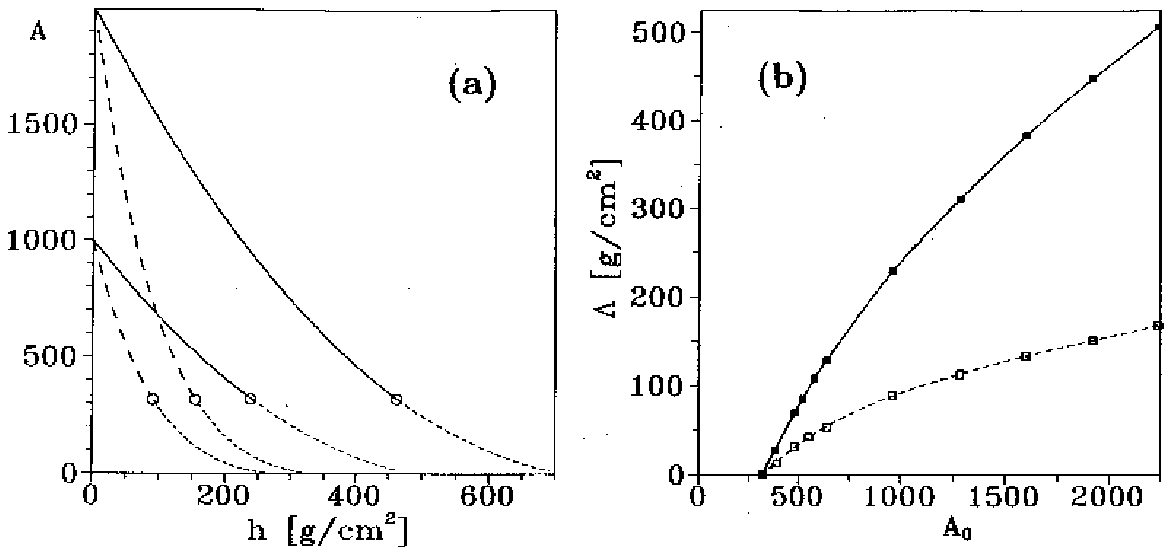}
\vskip -70pt
\begin{minipage}[t]{10cm}
\noindent \bf Fig. 2.  \rm
$(a)$ An example of the predicted decrease of actual size of
strangelet $A$ with depth $h$ of atmosphere traversed for two
different initial sizes: $A_0 = 1000~{\rm and}~2000$ and for two
scenarios of interaction: standard (solid and dotted lines) and 
tube-like (dashed and dotted lines). In both scenarios dotted lines
correspond to $A < A_{crit}$.
$(b)$ Atmospheric length $\Lambda$ after which initial strangelet
reaches its critical dimension, $A = A_{crit}$ ($=320$)
drawn as a function of
its initial mass number $A_0$ for standard
(solid line) and tube-like (dashed line) interaction. Consecutive
squares indicate points where (for $A_0>600$) 
$A_0/A_{crit} = k,
~(k=2,3,\dots )$.
\end{minipage}
\vskip 4truemm
In the numerical calculations we take $A_t = 14.5$ for the atmosphere
(assuming $25$\% of oxygen and $75$\% of nitrogen) and neglect the
influence of the interstellar gas. As a result one obtains the mass
number of strangelet registered at depth $h$, $A(h)$, as a function
of $h$ as shown in Fig. $2a$.  As one can see, bigger initial
strangelets (i.e., with higher mass number $A_0$) can penetrate much
more deeply into atmosphere untill $A(h)$ exceeds critical
$A_{crit}$, after which point they just evaporate by the emission of
neutrons. Notice much stronger degradation of the initial strangelets
for the case of the tube-like model interactions with air nuclei. As
we have already mentioned, our scheme displays two most extreme
situations in order to provide limits for all further possible
estimations. Because, for simplicity, we tacidly assume that air
nucleus destroys totally only the corresponding (equal to it) part of
the incoming strangelet, our estimation provides therefore only a
lower limit of what should be expected in more detailed calculations.
More refined estimations should take into account also the surface
tension effects and the possible dependence of $r_0$ and cross
section on the actual mass number of strangelet $A(h)$, which
translates immediately into dependence on $h$. However, their effect
would be of secondary importance only and at present experimental
situation such detailed and tedious estimations would be simply
premature.   

\vspace*{9cm}
\includegraphics{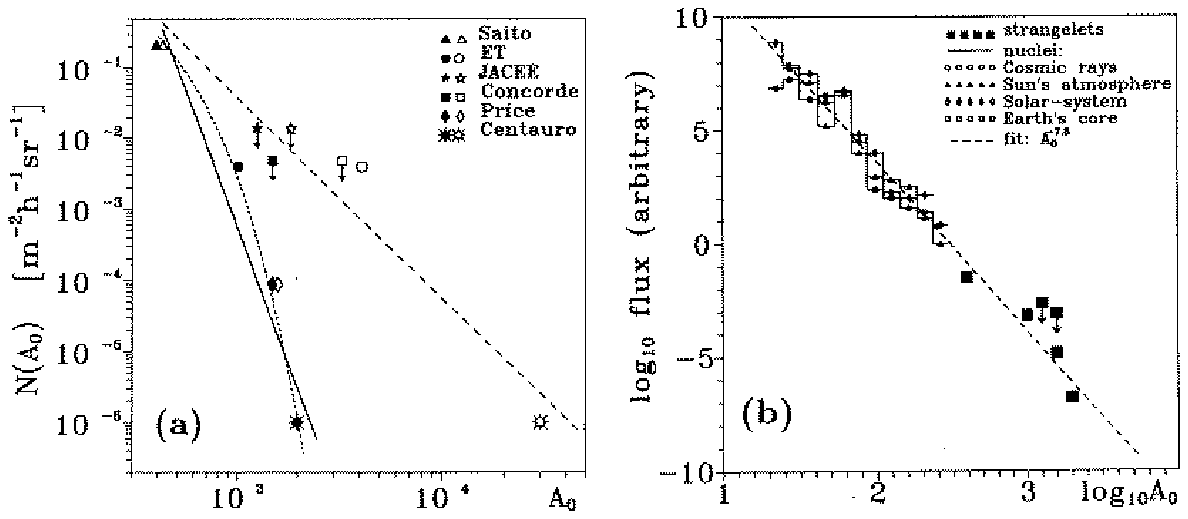}
\vskip -70pt
\begin{minipage}[t]{10cm}
\noindent \bf Fig. 3.  \rm
$(a)$ The estimation of the expected flux of strangelets 
on the border of atmosphere, $N(A_0)$, as a
function of their mass number
as obtained from standard (full symbols; solid line $\sim\,
A_0^{-7.5}$ and dotted line $\sim\, \exp \left( - A_0/130\right)$ )
and tube-like (empty symbols; dashed line $\sim A_0^{-2.8}$) 
models. The
respective points are taken from: {\it Saito} \cite{SAITO}, {\it ET}
\cite{S}, {\it Centauro} \cite{LATTES}, {\it JACEE} \cite{JACEE}, {\it
Concorde} \cite{CAP} and {\it Price} \cite{PRICE}. See text for
further details. $(b)$ Comparison of $N(A_0)$ from $(a)$ with
the abundance of elements in the Universe after \cite{Z}.
\end{minipage}
\vskip 4truemm
Suppose now that the energy per baryon in
strange matter is $\varepsilon = 919$ MeV and that number densities
corresponding to nuclear matter are $n = (110~{\rm MeV})^3$
\cite{FB}. In such situation, for $A\leq 1100,~E/A$ exceeds  already
$m_N$ but strangelet does not emit neutrons yet and starts to do so
only for $A\leq 320$, at which point $dE/dA$ exceeds $m_N$ \cite{FB}.
Below this limit strangelet decays rapidly by evaporating neutrons.
In view of these remarks it is remarkable that all possible
candidates for SQM have mass numbers near or slightly exceeding
$A_{crit}$, namely: $A = 350$ and $450$ in \cite{SAITO}, $A = 460$ in
\cite{ICH} and $A = 1000$ in \cite{S} and it is argued that Centauro
event contains probably $\sim 200$ baryons \cite{LATTES}.

Fig. $2b$ shows atmospheric length traversed after which strangelet
mass number $A$ becomes critical, $A = A_{crit}$ (starting from
different mass numbers $A_0$ of the initial strangelets). From this
figure one can read off that strangelets which are observed at depth
$200$ g/cm$^2$ should originate from strangelets of mass number $A_0
= 900$ at the top of the atmosphere whereas Centauro events observed
at the mountain altitudes would require original strangelet of $A_0 =
1800$. 

Let us now estimate flux of strangelets reaching our atmosphere
starting with the experimental data taken at different atmospheric
depths, cf. Fig. 3a. The experimental data {\it Saito, ET} and {\it
Centauro} on measured  fluxes on different atmospheric depths (which
can be interpreted in terms od SQM) are taken from \cite{SAITO},
\cite{S} and \cite{LATTES}, respectively. The corresponding upper
limits (no strangelets found so far), {\it JACEE} and {\it Concorde},
are from \cite{JACEE} and \cite{CAP}. Notice that Price data
\cite{PRICE} (assuming that what was observed was indeed a
strangelet) favour standard model. In terms of fits (for $3$ points:
{\it Saito, ET, Centauro}) one gets: $\sim\, A_0^{-7.5}$ for standard
model (full line, dotted line corresponds to $\sim\, \exp \left( -
A_0/130\right)$ in this model) and $\sim A_0^{-2.8}$ for tube-like
model (dashed line). The choice of the power-like or exponential form
for standard model was dictated by the analogy to nuclear
fragmentation and the expectation that decay (fragmentation) of a
strange star after its collision will result in the production of
strangelets with similar  distribution of mass \cite{NEMETH}.  It is
interesting to note that essentially the same exponent, $A_0^{-7.5}$,
as obtained in the standard model fit is observed also for occurence
of normal nuclei in the Universe \cite{Z}, cf. Fig. 3b.

\section{Summary and conclusions}

We summarize by stating that most probable the geometrical cross
sections of strangelets are not dramatically different from those for
the ordinary nuclear matter and cannot therefore explain their
apparent very high penetrability through the atmosphere. Instead we
propose to interprete such a penetrability of strangelets (already
discovered or to be yet observed) as indication of the existence of
very heavy lumps of SQM entering our atmosphere, which are then
decreased in size during their consecutive collisions with air nuclei
(i.e., their original mass number $A_0$ is reduced until $A =
A_{crit}$) and finally decay be the evaporation of neutrons. Assuming
that mass distribution of strangelets comprising this initial flux
follows power low $A_0^{\gamma}$ we estimate that $\gamma \geq 2$ is
consistent with present experimental data. Our result is consistent
with other estimations of the flux of strangelets at different
atmospheric depths. In particular it is interesting to note that the
best fit to data representing this dependence comes from our standard
model and it agrees very well with the existing data on chemical
composition of {\it normal nuclei} supporting therefore our
hypothesis that geometrical sizes of strangelets are not much
different from those of normal nuclei of the same $A$. 

\vskip 10pt
\noindent \large {\bf Acknowledgement}\normalsize
\vskip 10pt
\noindent
One of the Authors (G.W) is deeply gratefull to Profs. T. Cs\"org\"o,
P.Levai and J. Zimanyi for making his participation in {\it
Strangeness'96} conference possible and for its warm and extremaly
fruitful atmosphere.
 

\vfill\eject 

\begin{thebibliography}{99}\parindent=8truemm
\itemsep -1mm

\bibitem{SQM} E.Witten, {\sl Phys. Rev.} {\bf D30} (1984) 272. Cf.
also: Proc. of the {\it Int. Workshop on SQM in Physics and
Astrophysics}, eds. J.Madsen and P.Haensel, {\sl Nucl. Phys. {\bf B24}
(Proc. Suppl.)} (1991) and  Proc. of the {\it Int. Symp. on {\sc
Strangeness and Quark Matter}}, Kolymbari, Greece, Sept. 1-5, 1994,
World Scientific Pub. 1995.   

\bibitem{AO} C.Alcock and A.Olinto, {\sl Ann. Rev. Nucl. and Part.
Phys.} {\bf 38} (1988) 161.

\bibitem{FB} E.Farhi and R.L.Jaffe, {\sl Phys. Rev.} {\bf D30}
(1984) 2379; M.S.Berger and R.L.Jaffe, {\sl Phys. Rev.} {\bf C35}
(1987) 213.                 

\bibitem{ALOCK} C.Alcock and E.Farhi, {\sl Phys. Rev.} {\bf D32}
(1985) 1273; E.Farhi, {\sl Comm. Nucl. Part. Phys.} {\bf 16} (1986)
289.  

\bibitem{BAYM} G.Baym et al., {\sl Phys. Lett.} {\bf B160} (1985)
181; G.Shaw, G.Benford and D.Silverman, {\sl Phys. Lett.} {\bf 169}
(1986) 275. 

\bibitem{AFO} C.Alcock, E.Farhi and A.V.Olinto, {\sl Phys. Rev. Lett.}
{\bf 57} (1986) 2088.               

\bibitem{SEARCH} K.Borer et al., {\sl Phys. Rev. Lett.} {\bf 72}
(1994) 1415; D.Bevais et al. {\sl Phys. Rev. Lett.} {\bf 75} (1995)
3078. Also more dedicated experiment by J.Vandegriff et al., {\sl
Phys. Lett.} {\bf B365} (1996) 418, failed to observe strangelets in
terrestial atmosphere. For older propositions and results see
\cite{SQM}. 

\bibitem{SPIELES} C.Spieles et al., {\sl Phys. Rev. Lett.} {\bf 76} 
(1996) 1776 and references therein.

\bibitem{SAITO} T.Saito, Y.Hatano and Y.Fukada, {\sl Phys. Rev.
Lett.} {\bf 65} (1990) 2094.

\bibitem{PRICE} P.B.Price et al., {\sl Phys. Rev.} {\bf D18} (1978)
1382. 

\bibitem{ICH} M.Ichimura et al., {\sl Nuovo Cim.} {\bf A106} (1993)
843. 

\bibitem{JACEE} O.Miyamura et al., Proc. $24^{th}$ {\bf ICRC} 1, Rome
(1995) 890.

\bibitem{CAP} J.N.Capdevielle et al., Proc. $24^{th}$ {\bf ICRC} 1,
Rome (1995) 910.

\bibitem{LATTES} C.M.G.Lattes, {\sl Phys. Rep.} {\bf 65} (1980) 151.

\bibitem{HAENSEL} P.Haensel et al., {\sl Astron. Astrophys.} {\bf
160} (1986) 121 and {\sl Astrophys. Journal} {\bf 310} (1986) 261.  

\bibitem{PANAG} A.D.Panagiotou A.Petridis and M.Vassiliou, {\sl Phys.
Rev.} {\bf D45} (1992) 3134; M.N.Asprouli, A.D.Panagiotou and E.G\l
adysz-Dziadu\'s, {\sl Astropart. Phys.} {\bf 2} (1994) 167. 

\bibitem{KASUYA} M.Kasuya et al., {\sl Phys. Rev.} {\bf D47} (1993)
2153. 

\bibitem{S} T.Saito, Proc. $24^{th}$ {\bf ICRC} 1, Rome (1995) 898.

\bibitem{F2}  So far the results are negative: no evidence for
strangelets with $Z > 26 $ that survived passing of $\sim 100$
g/cm$^2$ of atmosphere was found.  

\bibitem{NEMETH} Cf., for example, percolation model of J.Nemeth et
al., {\sl Z.Phys.} {\bf A325} (1986) 347 and references therein.

\bibitem{Z} G.B.Zhdanov, {\sl Usp. Fiz. Nauk} {\bf 111} (1973) 109
[{\sl Sov. Phys. Uspekhi} {\bf 16} (1974) 642].

\end{thebibliography}
\end{document}